
 \magnification=\magstep1
\settabs 18 \columns
\hsize=16truecm

\def\s{\sigma}

\def\b{\bigskip}
\def\bb{\bigskip\bigskip}

\def\no{\noindent}
\def\r{\rightline}
\def\ce{\centerline}
\def\ve{\vfill\eject}

\def\r{\rightline}

\def\s{\sigma}

\def\harr#1#2{\smash{\mathop{\hbox to .25 in{\rightarrowfill}}
 \limits^{\scriptstyle#1}_{\scriptstyle#2}}}

\def\R{{\cal R}} 
\def\V{{\cal V}}

\def\today{\ifcase\month\or January\or February\or March\or April\or
May\or June\or July\or
August\or September\or October\or November\or  December\fi
\space\number\day, \number\year }

\r \today
\bb\bb\bb

\def\DD{\vec \bigtriangledown}


\def\Rrm{\hbox{\rm I\hskip -2pt R}}

\def\e{\rm e}
\def\d{\delta}
\def\p{\partial}

\def\sqr#1#2{{\vcenter{\vbox{\hrule height.#2pt
\hbox{\vrule width.#2pt height#2pt \kern#2pt
\vrule width.#2pt}
\hrule height.#2pt}}}}

 \def\1/2{{\scriptstyle{1\over 2}}}
 \def\a/2{{\scriptstyle{3\over 2}}}
 \def\5/2{{\scriptstyle{5\over 2}}}
 \def\7/2{{\scriptstyle{7\over 2}}}
 \def\3/4{{\scriptstyle{3\over 4}}}

\font\steptwo=cmb10 scaled\magstep2

\magnification=\magstep1

 \def\DD{\bigtriangledown}

\def\sqr#1#2{{\vcenter{\vbox{\hrule height.#2pt
\hbox{\vrule width.#2pt height#2pt \kern#2pt
\vrule width.#2pt}
\hrule height.#2pt}}}}

\def \r{\rightarrow}

\b

  {\ce {\steptwo   Heat and Gravitation. I. The  Action
Principle   }} 

\b 
  \ce{Christian Fr\o nsdal \footnote*{email: fronsdal@physics.ucla.edu}}
\b
  \ce{\it Physics Department, University of California, Los Angeles CA
  90095-1547 USA}
 \b

\def\sqr#1#2{{\vcenter{\vbox{\hrule height.#2pt
\hbox{\vrule width.#2pt height#2pt \kern#2pt
\vrule width.#2pt}
\hrule height.#2pt}}}}

\def \r{\rightarrow}

\b 
  \no ABSTRACT ~  Some features of hydro- and thermodynamics, as applied to
atmospheres and to stellar structures, are puzzling:  1.  The suggestion,
first made by Laplace, that our atmosphere has an adiabatic temperature
distribution, is confirmed for the lower layers, but the reason why it should
be so is difficult to understand.     2. The standard treatment of
relativistic thermodynamics does not favor a systematic treatment of
mixtures, such as the mixture of  a perfect gas with radiation. 3. The
concept of mass in applications of  general relativity to  stellar structures
is less than completely satisfactory.  4.   Arguments in which a concept of
energy plays a role, in the context of hydro-thermodynamical systems and
gravitation, are not always convincing.
  It is proposed that a formulation of thermodynamics as an action principle
may be a suitable approach to adopt for a new investigation of these
matters.   

This first article of a series formulates the thermodynamics of ideal
gases in a constant gravitational field in terms of an action principle
that is closely integrated with thermodynamics.  The
theory, in its simplest form, does not deviate from standard  practice, but
it lays the foundations for a more systematic approach to the various
extensions, such as the incorporation of radiation, the consideration of
mixtures and the integration with General Relativity.  We study the
interaction between an ideal gas and the photon gas, and propose a new
approach to this problem. We study the propagation of sound in a 
vertical, isothermal  column and are led to suggest
that the theory is incomplete, and to ask whether the true equilibrium state 
of an ideal gas may turn out  be adiabatic, in which case the role of
solar radiation is merely to compensate for the loss of energy  
by radiation into the cosmos.     An experiment with a
centrifuge  is proposed, to determine the influence of gravitation on
the equilibrium distribution with a very high degree of precision.

\b 

PACS Keywords: Atmosphere, photon gas, action principle.
  \ve
\no{\steptwo I. Introduction}

The premise of this paper is a conviction that the internal
consistency  of any physical theory is improved, if not assured, by a
formulation that is  based on a dynamical action principle. 
The consistency of a
set of equations is a delicate matter. When  a  modification
has to be made somewhere, to include an additional effect or another
degree of freedom, then  adjustments may have to be made
elsewhere, for consistency, but the need to do so is not
always evident. \footnote *{Whence the importance of the Onsager relations.} 
A   fully developed 
action principle treats all dynamical variables on an equal footing
and without constraints. Such a theory is said to be ``off shell",
something that has been crucial in the development of relativistic field
theories. When a change or a generalization is contemplated it
is done by modifying the lagrangian; changes in the equations of motion are
generated automatically and are believed to be mutually consistent. Another
advantage is that the equations of motion constitute a complete set of
equations that defines the theory; once the lagrangian is fixed, one need
not worry about the possibility that some important relation has been
overlooked, as is the case in applications of standard thermodynamics.  

Action principles have been used for some problems in 
hydro-thermodynamics, but as far as we know the injunction to avoid all
constraints has not been respected. Thus, in
hydrodynamics the temperature is related to the other variables by
what amounts to a constraint. The formalism has the aspect of a
partial projection of a complete dynamical theory on the space of
solutions.  

This paper is a study of atmospheres consisting of an ideal gas,
characterized by the ideal gas law and the expression for the
internal energy. A well known hydrodynamic action principle is
developed to include the temperature among the independent  
dynamical variables.  The theory is closely integrated with
thermodynamics and includes the specification of the Gibbs surfaces of
equilibrium configurations among th equations of motion.
 The hamiltonian is  identified with the thermodynamic function
$F + ST$, where $F(\V,T)$ is the free energy,
$S$ is the entropy and $\V$ is the volume. Variation of the lagrangian with
respect  to $T$ and $\V$ leads to the thermodynamic relations
$$
{\p F\over \p T}\big|_\V +S = 0, ~~{\p F\over \p\V}\big|_T +p = 0.
$$
In the case of an ideal gas, this equation 
takes the form of the polytropic relation $\V T^n $ = constant. On
shell; that is, by virtue of the first of these relations, the hamiltonian
density reduces to  the familiar formula for the internal energy,
$U = n\R   T$.

It is important to ask to what extent the observed polytropic
relations are to be attributed to intrinsic properties of the gas, or to
radiation. Although the question is somewhat academic, since it  
does not directly affect the main applications, it is natural to ask what 
are the  natural configurations of an isolated atmosphere, one that is
not exposed to radiation. Our
understanding of atmospheres will be incomplete without an answer
to this question.

 The statement that any two thermodynamic systems, each in a state of
equilibrium with a well defined temperature, 
and in thermal equilibrium with
each other, must have the same
temperature, is a central tenet of thermodynamics. 
A natural generalization is
that the temperature, in an extended but closed system in a state of
equilibrium,  must be uniform, and
there is near  universal agreement that  this remains true in the
presence of gravitational fields. This is important for the
understanding of 
terrestrial and stellar atmospheres, where the gravitational forces 
create a
non-uniform density distribution.

It must be stressed that an actual, terrestrial or stellar, atmosphere is
not isolated and is not a thermodynamic state of equilibrium. It is, at
best, a stationary configuration in which the effects of incoming and
outgoing radiation are balanced in what is called radiative equilibrium.
When such atmospheres are described by a mathematical model, then this
quiescent state is mapped to the equilibrium of the mathematical model. 
What is puzzling, is that the
mathematical model, intended to describe an isolated
gas, with the gravitational field included in standard fashion, turns out
to be a model (in fact, the standard model) of an actual atmosphere that
is far from being isolated. This mystery has an easy, but unpopular 
resolution: If one could concede that the true equilibrium state may be
isentropic instead of isothermal, then the role of incoming radiation (in
the case of the earthly atmosphere), or internal generation of energy (in 
the case of stars) would be simply to make up for the loss of energy  
radiated into the cosmos. So far, we have not found an attractive
alternative.
\b

\ce{\bf Outline}

Hydrodynamics is a theory of continuous
distributions of matter,  described in the simplest case by two 
fields or
distributions: a density field and a velocity field, both defined over
$\Rrm^3$ or a portion thereof.  The role of temperature is often 
constrained, as it is
taken to be determined by the density and the pressure.
Classical thermodynamics, on the other hand, is   the study of
states of equilibrium, with uniform density and temperature,  
and relations
between such states.  In this
context, extremum principles first formulated by Gibbs (1878) play an
important role;  see   for example 
 Callen  (1960), but the extension
of thermodynamics to systems in which the dynamical 
variables are fields
on $\Rrm^3$ is not immediate and in fact  variational
principles are seldom invoked in studies of such systems. 
Investigations
that deal with  flow of matter or with temperatures 
that vary in time and
space are found under the heading of heat transfer, 
fluctuations,
thermodynamics of irreversible processes and radiation hydrodynamics. 
  See for example Stanyukovich (1960), 
 Castor (2004), M\"uller (2007).

In this introductory section we study a simple system from the point of 
view of hydrodynamics, on the basis of a well known action principle that
incorporates the continuity equation and Bernoulli's equation.  The
temperature is not an independent variable but is assumed to be 
given by the
ideal gas law. We stress the role of mass (Section I.5). The  potential is
chosen appropriately  for an atmosphere that is either isothermal or
polytropic.  
  In
applications to the earthly atmospheres, and to stellar structure as
well,  the polytropic model is  universally preferred. \footnote *{A
book of more than 700 pages, devoted entirely to polytropes, has
appeared as recently as 2004. (Horedt 2004)}   Section I.6 is a short
introduction to polytropic atmospheres. It is remarkable that this theory
already predicts the observed  temperature lapse rate for our atmosphere.
The only   parameters are the known values of the adiabatic index and the
effective atomic weight of the gas. There is no parameter that can be
interpreted as a measure of the influence of radiation and no indication
within the theory that the  (radiative) equilibrium becomes isothermal in the
limit when the radiation is turned off. We offer a brief review of the
history of the  polytropic atmosphere (Section I.7). 
 
In Section II  we take the action
principle off shell, to include the temperature as   an independent field
variable (Sections II.2-5).    Variation of the lagrangian with respect to
the temperature is related to a standard, thermodynamical relation, $\p F
/\p T = -S$.  The potential is refined to give the correct expression for
the internal energy on shell; the off shell hamiltonian  appears to be
new.

In Section II.6 we include interactions with the photon gas,
by adding the Stefan-Boltzmann energy to the hamiltonian. The
radiation pressure appears as an addition to the thermodynamic pressure.
The effect is to increase the effective polytropic index (towards the
upper limmit of 3) at high temperatures. No approximation of the kind 
$p_{\rm gas}/p_{\rm total} = $ constant, used by Eddington, is needed.

When this approach is compared to modern hydro-thermodynamics, 
restricted to
adiabatic processes, we find  that, so long as the radiation pressure is
small in comparison with the gas pressure,  all the basic equations are
identical.   But the incorporation
of additional features is much more straightforward within the variational
formulation.  

Section II.7   examines the intimate integration of this action
principle with thermodynamics. It is shown that the  
action  principle, restricted to stationary configurations, 
is in full accord with the  extremal conditions formulated by Gibbs,  
for energy and for entropy. When the (localized) internal energy is
interpreted as a hamiltonian one recovers all the equations of motion
derived from the action principle. It is pointed out that Gibbs' axiom of
extreme entropy, which is not a result of the dynamical action principle,
when applied to an isoloated  continuous system, demands uniform entropy
density  rather than uniform temperature.

 There is an extensive literature on the propagation of shock waves in
a polytropic atmosphere but, as  far as we know, none that deals with
the vertical propagation of a disturbance (sound waves or shock waves) 
in an isothermal gas in a gravitational field. We study this problem
in Section II.8. The result is unexpected and provokes further
examination of the standard theory; we  argue that it
 is incomplete. The injunction that  equilibria
(and some other configurations!) must  be isothermal does
not  seem to be a special case of a more general doctrine; there is a 
striking lack of continuity.

 The equilibrium state of any ideal gas with
a finite  adiabatic  index is essentially polytropic. Here it is important
to make precise what we mean by equilibrium. We use the term in the
context of
 a mathematical model; it refers to a solution of the equations
of motion with the property that the flow vanishes and all the
fields are time independent. Polytropic models of  earthly and
stellar atmospheres are very widely used, and the stationary
configurations of such atmospheres are equilibria in this sense,
though they are not states of true, thermodynamic equilibrium. Since
the density is not uniform, neither  is  the  temperature. An
issue that we wish to discuss is the precise role that is played
by radiation. We should hope to develop an understanding of what would
happen if the intensity of radiation were continuously reduced to zero.
The possibility that the limit might  turn out to be other than isothermal
is  not easy to accept,  for it  goes against one of the basic tenets of
thermodynamics: Clausius'  statement  of the second law (Section II.9).
The question is not entirely academic, but it has no direct bearing on the 
validity of our approach, for we apply it to the standard, 
polytropic atmospheres, and to problems where gravitation does not
intervene.

It is important  to incorporate the isothermal
equilibrium into a dynamical framework; that is, to
modify or extend the standard theory of polytropic atmospheres, to
include a parameter or variable to represent the intensity of radiation
and that would allow the effect of radiation to be reduced to zero,
resulting in an isothermal equilibrium in the limit.  In the event that
experiment should validate the isothermal   atmosphere the need to
construct such a theory would become urgent. (Section II.9)

There seems to be a dearth of experimental data.  
We study an ideal gas in a centrifuge and invoke the equivalence principle
to relate this situation to  atmospheres.   Experiments are proposed.
(Section II.10).

Some further speculations are in the Appendix.

\b

\ce{\bf Applications to astrophysics.}

The simplest form of the action principle studied here is the non
relativistic approximation to a fully relativistic theory that, with the
addition of the Einstein-Hilbert action for the metric, has been applied
to the dynamics of certain stars (Fronsdal 2007, 2008). The results are
close to those of the traditional approach except for certain 
features that make the present  approach more attractive to us, such as
the preservation of the non-relativistic equation of continuity. The
present study was undertaken as preparation for an attempt to take into
account the radiation field. In this paper the radiation energy
(Stefan-Boltzmann energy) is included in the lagrangian; this
automatically generates the radiative correction to the pressure 
(Section II.6); in a future paper we plan to lift it to a relativistic
version and return to the study of stellar dynamics.

\b \b
\ce{\bf I.1. Hydrodynamics}

Basic hydrodynamics deals with a density field
$\rho$ and a velocity field $\vec v$ over $\Rrm^3$, subject to two
fundamental equations, the equation of continuity,
$$
\dot \rho + {\rm div}(\rho \vec v) = 0,~~ \dot \rho := {\p\rho\over \p
t},\eqno(1.1)
$$
and the hydrodynamical equation (Bernoulli 1738)
$$
-{\rm grad}~ p = \rho{D\over Dt}\vec v := \rho(\dot{\vec v} + \vec v\cdot
{\rm grad} ~\vec v).\eqno(1.2)
$$
This involves another  field, the scalar field $p$, interpreted as the
local pressure. The theory   requires an
additional equation relating $p$ to $\rho$. It is always
assumed that this relation is local, giving $p(x)$ in terms of the
density (and the temperature) at the same point $x$, and instantaneous.

\ve

\ce{\bf I.2. Irrotational flow}

Since there is enough to do without taking  on  difficult problems of
turbulence, we shall  assume, here and throughout, that the velocity field can be
represented as the gradient of a scalar field,
$$
\vec v = -{\rm grad}~ \Phi.\eqno(1.3)
$$
In this case the hydrodynamical condition (1.2) is reduced to
$$
   {\rm grad} ~p = \rho~{\rm grad}~ (\dot\Phi - \vec
v^2/2).\eqno(1.4)
$$
To complete this system one still needs a relation between the fields $p$
and
$\rho$.

Assume that there is a local
functional
$V[\rho]$ such that
$$
p = \rho {\p V\over \p \rho}- V.\eqno(1.5)
$$
In this case
$$
{\rm grad} \,p = \rho {\p V\over \p \rho}\eqno(1.6) 
$$
 and the equation (1.4) becomes, if $\rho
\neq 0$,
$$
{\rm grad}~ {\p V\over \p \rho}= {\rm grad}~(\dot \Phi -   \vec v^2/2)\eqno 
$$
or
$$
{\p V\over \p \rho}=  \dot \Phi -  \vec v^2/2 + \lambda,~~ \lambda ~
{\rm constant}.\eqno(1.7)
$$
  The potential $V[\rho]$ is defined by $p$
modulo a linear term, so that the appearance of an arbitrary constant
is natural. It will serve as a
Lagrange multiplier.

{\it The introduction of a velocity potential guarantees the existence of
a first integral of the motion,
a conserved energy functional that will play an important role in the
theory.}

It will turn out that $V$, with the inclusion of a term linear in $\rho$ that remains
undetermined at this stage,  is related to the internal energy density.
 
\b\b
\ce{\bf I.3. Variational formulation}

Having restricted our scope, to account for irrotational flows only,  we have
   reduced the fundamental equations of simple hydrodynamics to the
following two equations,
$$\eqalign{ 
\p_\mu J^\mu& =  0,~~ J^t := \rho,~ \vec J := \rho \vec v, \cr &
\p V/\p \rho =  \dot \Phi -  \vec v^2/2 + \lambda,\cr}\eqno(1.8)
$$
together with the defining equations
$$
\vec v = -{\rm grad}~ \Phi,~~ p := \rho V' - V.\eqno(1.9)
$$
It is well known that these equations are the Euler-Lagrange
equations associated with the action (Fetter and Walecka  1980)
$$
A[\rho,\Phi] = \int dtd^3x ~{\cal L},~~ {\cal L} =  \rho(\dot\Phi - \vec v^2/2 +
\lambda) - V[\rho].\eqno(1.11)
$$
 The value of this last circumstance lies in the fact that the
variational principle is a better starting point for generalizations,
including the incorporation of symmetries, of special relativity, and the
inclusion of electromagnetic and gravitational interactions. It also gives us a
valid concept of a total energy functional.
 
\b\b 

\ce{\bf I.4. On shell relations }

   The
action (1.10) contains only the fields $\Phi$ and $\rho$. The
Euler-Lagrange equations define a complete dynamical framework, but only
after specification of the functional $V[\rho]$. The pressure was
defined by Eq.(1.9), $p := \rho V' - V$, and one easily verifies that,
  by virtue of the equations of motion,
$$
p = {\cal L} ~~({\rm on~ shell}). 
$$
This fact has been noted, and has led to the
suggestion that the action principle amount to  minimization of $\int
p$ with respect to variations of $p$ defined by thermodynamics
(Taub 1954), (Bardeen  1970), (Schutz 1970).
   But  an off shell action is needed.  
 The lagrangian density  is not a
thermodynamic function, since it depends on the time derivatives of the
variables. After adopting the action (1.10) it remains  to  relate the
choice of the potential $V$ to the thermodynamical properties of the
fluid. We shall find that the properties that define an ideal gas lead to
a unique expression for $V$. 

It is useful to reflect on the meaning of Eq.(1.6) as well.    In a more
general situation, in which the potential $V$ depends on the temperature, we
would have, instead of (1.6), the identity
$$
 \rho~ {\rm grad}~ {\p V\over
\p \rho} ={\rm grad}~ (\rho {\p V\over \p \rho}-V) + \rho{\p V\over \p
T}\,{\rm grad }\,T.\eqno(1.11)
$$
The expression (1.5) for the pressure would be valid ``on shell" if  
 the action principle includes variations of the
temperature as an independent variable, giving the on shell condition  $\p V/\p T = 0$.

 There is a unique thermodynamical function that is a prime candidate for being identified with 
 the  potential $V$.
 It is the function
 $$
 f(\V,T)+sT 
 $$
of three independent variables,  
 where $f$ and $s$ are the free energy density and the entropy density.
The integration with thermodynamics is explored in Section II.7.

\ve
 
\ce{\bf I.5. The mass}

To speak of a definite, isolated physical system we must fix some
attributes, and among such defining properties we 
include the mass. We insist on this as it shall turn out to be crucial to
the stability of stellar atmospheres (Fronsdal 2008). The density $\rho$
will be taken to have the interpretation of mass density, and the total
mass is the constant of the motion
$$
M = \int d^3 x ~\rho.
$$
Such integrals, with no limits indicated, are over the domain $\Sigma$ of
definition of $\rho$, the total extension of our system in \Rrm$^3$.

Since the total mass is a constant of the motion it is natural to fix
it in advance and to vary the action subject to the constraint
$\int_\Sigma d^3x\, \rho(x) = M$. The parameter $\lambda$ takes on the role of  a
Lagrange multiplier and the action takes the form
$$
A = \int _\Sigma d^3x\Big(\rho(\dot\Phi - \vec v^2/2) -V\Big)
+\lambda\Big(\int _\Sigma d^3x \rho - M\Big).
$$

The conservation of mass has important implications for boundary 
conditions. 
\b\b
 
\ce{\bf I.6. Equation of state and equation  of change}

An ideal gas at equilibrium, with constant temperature, obeys the gas law
$$
p/\rho = \R T.\eqno(1.12)
$$
Pressure and density are in cgs units and 
$$
\R = (1/\mu) \times .8314\times 10^8~erg/K ,
$$
 where $\mu$ is the 
molecular weight. The gas law
is assumed to hold, locally at each point of the gas.
Effective values of $\mu$ are
$$
{\rm Atomic ~hydrogen:}~\mu = 1,~~ {\rm Air:}~\mu = 29,~ ~{\rm Sun:}~ \mu = 2.
$$

Equation (1.12) is the only equation that will be referred to as an `equation of
state'.
 Other relations, to be discussed next, are `equations of
change', this term taken from Emden's ``Zustands\"anderung", for their
meaning is of an entirely different sort. Most important is the polytropic
relation
$$
p = A\rho^{\gamma'}, ~~ A, \gamma' ~{\rm constant}.\eqno(1.13)
$$
This relation defines a polytropic \underbar {path} or \underbar{polytrope} in 
the $p,\V$ diagram ($ \V = 1/ \rho $). A polytropic atmosphere is one in which, 
as one moves through the gas, the
variables $\rho$ and $p$ change so as to remain always on the same polytrope.
Eq.(1.13) is a statement about the system, not about the gas {\it per se}.
The validity of (1.13) for an actual atmosphere cannot be inferred from
the early laboratory experiments.

The index of the polytrope is the positive number $n'$ defined by
$$
\gamma' =: 1 + {1\over n'}.
$$
Important special cases are
$$
n' = 0,~~ \gamma' = \infty,~~\rho = {\rm constant},
$$
$$
\gamma' = C_P/C_V, {\rm ~specific ~entropy~ = constant},
$$
$$
n' = \infty, ~~\gamma' = 1,~~T = {\rm constant}.
$$
 Numbers $\gamma,~n$ are defined by
$$
\gamma := C_P/C_V =: 1 + {1\over n}.
$$
 
The number $n$ is the
adiabatic index of the gas. According to statistal mechanics $2n$ is the number
of degrees of freedom of each molecule in the gas.   That atmospheres tend to be
polytropic is an empirical fact.   

The case that
$\gamma' =
\gamma$ is of a special significance. 
A polytrope  with $\gamma' = \gamma$ is a path of  constant  specific
entropy; changes along such polytropes are reversible and adiabatic; these
polytropes and no others are adiabats.

Fix the constants $A, \gamma' $ in (1.13) and consider an associated  stationary,
polytropic  atmosphere.   If both (1.12) and (1.13) hold we have (Poisson
1835)
$$
p = {\rm const.} \,\rho^{\gamma'},~~p = 
{\rm const.} \,T^{ {\gamma'\over \gamma'-1} },~~ T = {\rm const.}\,
p^{1-1/\gamma'}.\eqno(1.14)
$$
In any displacement along a polytrope from a point with pressure $p$ and 
temperature $T$, we shall have 
$d\rho/\rho = (1/\gamma')dp/p$, so that an increase in pressure leads to 
an increase in density that is greater for a smaller value of $\gamma'$.  If
a parcel of gas in this atmosphere is pushed down to a region of higher
pressure, by a reversible process, then it will adjust to the ambient pressure.
If $\gamma  > \gamma'$, then it will acquire a density that is
lower than the environment; it will then rise back up; this atmosphere is
stable. But if $\gamma' > \gamma$ then the parcel will be denser than the
environment and it will sink further; this atmosphere
is unstable to convection. Thus we have:
\b
{\it A stable,  polytropic atmosphere must have $\gamma' < \gamma,~~ n' > n$.}
\b
\no Most stable  is the isothermal atmosphere, $\gamma' = 1$.
\b
In hydrodynamics,    the isothermal atmosphere can be given a lagrangian 
treatment by taking
$$
V = \R T \rho \ln\rho.\eqno(1.15) 
$$
 We
suppose that the gas is confined to  the section  $z_0<z<z_0 + h$ of a
vertical cylinder with base area ${\cal A}$ and expect the density to fall off
at higher altitudes.  A plausible action density, for a
perfect gas at constant temperature $T$ in a constant gravitational field $\phi
= gz$, $g$ constant, is
$$
{\cal L}[\Phi,\rho] = \rho\,(\dot\Phi - \vec v^2/2  - gz + \lambda) -
{\cal R}T\rho\ln\rho.\eqno(1.16)
$$
We may consider this an isolated system with fixed mass and fixed
extension.

At equilibrium $\dot\Phi = 0, \vec v = 0, \dot \rho = 0$ and the
equation of motion is
$ V' = {\cal R}T(1 + \ln\rho) = \lambda - gz,$
hence
$$
\rho(x,y,z) = \e^{ -1+\lambda/\R T}\e^{-gz/\R T},~~ M = {\cal A}{ {\R}
T\over g}\e^{-1 +\lambda/\R T}(1- \e^{-gh/RT})~\e^{-gz_0/\R T}
$$
and after elimination of $\lambda$
$$
\rho = {gM\over {\cal A}{\R} T}{\e^{-g(z-z_0)/\R T}\over 1-
\e^{-gh/RT} },~~ p={gM\over {\cal A}}~{\e^{-g(z-z_0)/\R T}\over 1-
\e^{-gh/RT}}.\eqno(1.17)
$$
There is no difficulty in taking the limit
$h\rightarrow \infty$. The volume becomes infinite but it can be
replaced as
a variable by the parameter $z_0$. This atmosphere is stable.

  The isothermal atmosphere
is usually abandoned in favor of the polytropic atmosphere.
\b
A polytropic gas can be described by the lagrangian (1.10), with
$$
V = \hat  a\rho^{\gamma'}, ~~ \hat a, \gamma' ~{\rm constant}.
$$
Variation with respect to $\rho$  gives
$$
p ={\hat a\over n'}\rho^{\gamma'},~~{1\over n'} = \gamma'-1.
$$
The temperature does not appear explicitly 
but is taken to be determined by the
gas law, $p = \R\rho T$. Among the many
applications  the following are perhaps the most important. In the case of sound
propagation the gas is initially awakened from equilibrial turpor and then left
in an isolated, frenzied state of oscillating density and pressure,
with the temperature keeping pace in obedience to the gas law
(Laplace 1825, Pierce 2008). All three of the relations (1.14) are
believed to hold, with
$\gamma' =
\gamma$. The oscillations are usually  too rapid for
the heat to disseminate and equalize the temperature,
so that the neglect of heat transfer
is justified. It should be emphasized, however, that these rapid variations of
the temperature with time are not predicted by the theory; Laplace's
postulate that $(  {\delta\dot\rho}/\delta\rho)/( {\delta \dot T}/\delta T)  =
n'$ is bold, independent assumption.

In applications to atmospheres one often postulates the
polytropic equation of change (1.13) and obtains the temperature from 
the gas law. Understanding the resultant temperature gradient in terms of
convection, or as the effect of the heating of the air by solar radiation,
or both, is one of the main issues on which we have hoped to gain some
understanding.

At mechanical equilibrium $\vec v = 0, \dot \rho =
0$ and $\lambda - gz= \hat a\gamma\rho^{1/n}$, hence
$$
  \rho =
({\lambda - gz\over \hat a\gamma})^n.  
$$
Since the density must be
positive one does not fix the volume but assumes that the atmosphere
ends at the point $z_1 = \lambda/g$. Then
$$
 M= {{\cal A}}({g\over
\hat a\gamma })^n\int_{z_0}^{z_1}(z_1-z)^ndz = {{\cal A}h\over
n+1}({gh\over \hat a\gamma})^n. \eqno 
$$
This fixes  $h$ and thus $z_1$
and $ \lambda$. If the atmosphere is an ideal gas then 
 the temperature varies with altitude according
to
$$
{\cal R}T = p/\rho = {\hat a\over n}\rho^{1/n} = g{ z_1-z\over n+1}.
\eqno(1.18) 
$$ 
Because the lagrangian does not contain $T$ as a dynamical variable it is
possible to impose this condition by hand.  

 One would not apply this theory
down to the absolute zero of temperature, but even without going to  
extremes it seems risky to  predict  the temperature of
the   atmosphere  without having made any explicit assumptions about the
absorption or generation of heat that is  required to sustain it.
Yet this has been the basis for the phenomenology of stellar structure, as well
as the earth's atmosphere, from the beginning (Lane 1870, Ritter 1878). 

{\it The success of the polytropic model is amazing,   but the theory is
incomplete since it does not  account for heat flow, nor convection, both of
which are needed to complete the picture.}

For air, with
molecular+ weight 29, $\R = 2.87\times 10^6 ergs/gK$ and $n = 2.5.$  At
sea level,\break$g = 980 cm/sec^2$,  the density  is   $\rho  =1.2
\times 10^{-3}  g/cm^3$, the pressure $ p =1.013 \times 10^6
dyn/cm^2$.  Thus
$$
p/\rho = .844 \times 10^9 cm^2/sec^2, ~~ T = T_0
= 294K,~~ z_1 = 3.014\times 10^6 cm \approx 30km,
$$
and the dry lapse
rate at low altitudes is  $ -T'=294/ z_1= 9.75  K/km.$ The opacity that is
implied by this is mainly due to the presence of $CO_2$ in the atmosphere.
Humidity increases the opacity and decreases the lapse rate by as much as a
factor of 2. (A temperature difference of
70 degrees over 12 000 m was observed on a recent flight over Europe.)

 The specific internal energy of this hydrodynamical model is 
 $np/\rho = n\R T$, as it
 should be for an ideal gas.

\ve
 \ce{\bf I.7. Historical notes on polytropic atmosphere}

 Observations of reversible transformations of
near-ideal gases,  carried out during the 19th century, can be summarized in
what is sometimes called the laws of Poisson, 
 $$
 \rho\propto T^{n'},~~ p \propto T^{n'+1},~~p \propto
 \rho^{\gamma'},~~ 
 \gamma' = 1 + {1\over n'} ~{\rm constant}.
 $$
In the original context all the variables are constant and uniform. 
The exponents as well as the coefficients of proportionality 
are the same for
all states that are related by reversible transformations. 
Statistical mechanics
explained this result and confirmed the experimental value 
$\gamma' = \gamma  = C_P/C_\V$. As far as can be ascertained, the presence of 
terrestrial gravitation
and ambient radiation had no effect on these experiments. In a first
extrapolation the same relations were taken to hold locally in
dynamic situations, as in the case of sound propagation.  
{\it The gas is not in thermal equilibrium and the variation of the
temperature from point to point, and with time, is obtained from the gas
law. This extension of an important thermodynamical relation to the case
of a nonuniform system is taken for granted.}

For the atmosphere of the earth it was at 
first proposed that the temperature 
would be uniform. However, the existence
of a temperature gradient was soon accepted as an incontrovertible 
experimental
fact. The first recorded recognition of this, together with an attempt at
explaining the same, may be that of  Sidi Carnot, in the paper in which
he created the science of thermodynamics (Carnot 1824). Carnot quotes Laplace: 
``N'est-ce pas au refroidissement  de l'air par la dilatation qu'il faut
attribuer le froid des r\'egions superieures de l'atmosphere? Les raisons
donn\'ees jusqu'ici pour expliquer ce froid  sont tout a fait insuffisantes; on
dit  que l'air des r\'egions elev\'ees, recevant peu de chaleur reflechie par la
terre, et rayonnant lui meme vers les espaces celestes, devait perdre de
calorique, et que c'etait l\'a la cause de son refroidissement; ... " This may be
the first time that  the influence of radiation is
invoked. The temperature gradient is attributed to the greenhouse effect, 
and Laplace was an early skeptic, for he continues   ``...mais cette explication
ce trouve detruite si l'on remarque qu'a
\'egale hauteur le froid regne aussi bien et meme avec plus d'intensit\'e sur
les plaines elev\'ees que sur les sommets des montagnes ou que dans les parties
d'atmosphere \'eloignees du sol." It is not clear that the two explanations are
at odds with each other; Laplace apparently postulates that the atmospheres over
lands at different elevations are related by adiabatic transformations,
but without explaining why.

By rejecting the role of radiation as the cause of the temperature gradient, 
Laplace seems to suggest that the same would be observed in an atmosphere
subject to gravitation but totally isolated from radiation, neither exposed to
the radiation coming from the sun nor radiating outwards. As was strongly
emphasized in   later phases of this debate, this would contradict the belief
that the thermal equilibrium of any isolated system, gravitation and other
external forces notwithstanding, is characterized by a uniform temperature. 

In 1862 W. Thomson, in the paper ``On the convective equilibrium of the 
temperature in the atmosphere", defines convective equilibrium with these words
``When all parts of a fluid are freely interchanged and not sensibly influenced
by radiation and conduction, the temperature is said to be in a state of
convective equilibrium." He then goes on to say that an atmosphere that is in
convective equilibrium is a polytrope, and we think that he means an adiabat,
because of the words ``freely interchanged", although 
the value of the polytropic index is taken from experiment and not from
statistical mechanics. At first sight the clause ``and not sensibly influenced
by radiation" would seem to imply that his remarks  apply to an
isolated atmosphere,  indicating that   a temperature gradient would
persist in the absence of radiation,  but this conclusion would be
premature, as we shall see.\break

Later, Lord Kelvin had doubts about what he called the 
Boltzmann-Maxwell doctrine 
  and especially its application to the isolated atmosphere. See his
Baltimore lectures,  onwards(Kelvin 1904). \footnote * {``The time integral
of the kinetic energy of any atom will be equal to the time integral of the
kinetic energy of any other atom. This truism is simply and solely all
that the Boltzmann-Maxwell doctrine asserts for a vertical column of a
homogeneous monatomic gas."  }

In 1870 H.J. Lane  made the bold assumption that the laws of Poisson may be 
satisfied in the Sun. The terrestrial atmosphere (or part of it) had already been
found  to be well represented by the same relations. Referring to Lane's
paper  Thomson, now Lord Kelvin, explains how convective equilibrium
comes about (Thomson 1907). He argues that the atmosphere is not, cannot be, at
rest, and this time radiation plays an essential role. The  upper layers loose
heat by radiation and the lower temperature leads to  an increase in density.
This produces a downward current that mixes with a compensating upward drift of
warmer air. This continuing mixing takes place on a time scale that is too short
for adjacent currents to exchange a significant amount of heat by conduction or
radiation, especially since the variations of temperature are very small.   It
is evident that Thomson offers his explanation of the temperature gradient to
account for its absence in an isolated atmosphere,
for he says that, ``an ideal atmosphere, perfectly isolated from absorption as
well as emission of radiation, will, after enough time has passed, reach a state
of uniform temperature, irrespective of the presence of the gravitational
field".   Thomson accepts the mechanism of  Laplace and Carnot, as it is at work
in the real atmosphere, but he goes further. 
 He believes that the lower temperature aloft is intimately tied to the
existence of radiation,  implying that it is driven by net outwards radiation.
(The effect of solar radiation on the terrestrial atmosphere is not explicitly
mentioned.)  It is difficult to judge whether or not Thomson is in disagreement
with Laplace, but the precision of his statements represents a marked
improvement over his predecessors and his earlier work.   

The principal developers of the field,
Ritter (1878-1883) and Emden (1907), seem to accept the idea of convective
equilibrium. It may be pointed out, however, that this mechanism is in no way
expressed by the equations that these and other authors use to predict the
behaviour of real atmospheres. {\it The concept of convective equilibrium is
introduced to 
 one purpose only: to avoid contradiction with firmly established belief in the  
isothermal equilibrium of isolated systems. It receives no quantitative
theoretical treatment.}

Nor was it accepted by everybody. A famous incidence involves Loschmidt (1876),
who believed that an isolated atmosphere, at equilibrium in a gravitational
field, would have a temperature gradient.  But  arguments presented by Maxwell
and Boltzmann (1896) led Loschmidt to withdraw his objections, which is hardly
surprising given the authority of these two. Nevertheless, it may be pointed out
that no attempt was made, to our knowledge, to settle the question
experimentally, until recently    (Graeff 2008). 

An alternative to convective equilibrium was proposed by Schwarzschild (1906)
and  critically examined by Emden.  To understand how it  works we turn to
Emden's book of 1907, beginning on page 320. Here he invokes a concept that is
conspicuously absent from all his calculations on polytropic spheres in the rest
of the book: heat flow. He posits that the atmosphere is not
completely transparent, and that heat flow is an inevitable consequence of the
existence of a temperature gradient. {\it The most important observation is that
heat flow is possible in stationary configurations ($\dot T = 0$) provided that
the  temperature gradient is constant.} We take this to be an implicit
reference to the heat equation, the first such reference in the book(!). 
The heat flux due to conduction and radiation is usually thought to be
expressed as  
$$
\vec F =  - C\vec \bigtriangledown T,~~F^i = -C^{ij}\p_j T,
$$
where the tensor $C$ includes the thermal conductivity as well as the effective
coefficient of heat transfer by radiation. The divergence of the flux is the
time rate of change of the temperature due to conduction and radiation. In a
stationary, terrestrial atmosphere, with no local energy creation, this must
vanish. Emden's atmospheres are polytropes,
with temperature gradients that are constant. It appears that he takes
$C$ to be uniform.
 That is surprising, and interesting, for he suggests that the entire
edifice
 implicitly demands that this condition, of a constant
heat flow, must be satisfied.

We note that the direction of flow is from hot to cold, outwards.  In applications to
 planetary atmospheres, with no local energy generation, this calls
for an explanation, since the ultimate source of energy is above. Here we have
to return to the oldest explanation of the existence of a temperature gradient,
dismissed by Laplace ({\it op. cit.}): the greenhouse effect.  The atmosphere is
highly transparent to the (high frequency) radiation from the Sun  but opaque to
the thermal radiation to which it is converted by the ground. The atmosphere is
thus heated from below! 

If the atmosphere is stable in the sense discussed above, when $\gamma'
\leq C_P/C_V$, then it is not necessary to assume that any convection takes
place. In this case one speaks of (stable) `radiative equilibrium'. Convective
equilibrium may step in when the stationary atmosphere is unstable, but
it is no longer used to explain the existence of a
temperature gradient.  

A difficulty is present in all accounts of stellar structure up to  1920. 
The energy observed to be emitted by the Sun, attributed to 
contraction of the mass and the concomitant release of internal energy, was far
too small to account for the age of the sun as indicated by the geological
record. The situation changed with the discovery of thermonuclear energy
generation. Now there is plenty of energy available. At the same time there
arose the realization that convection sometimes plays a very modest role; the
concept of convective equilibrium was put aside and with it, Kelvin's explanation
of the temperature gradient. According to Eddington (1926), who is more
concerned with stars than with our atmosphere,  ``convective equilibrium"
must be replaced by ``radiative equilibrium" in the sense of
Schwarzschild. He does not claim that this new concept accounts for the
temperature gradient as well as Kelvin's convective equilibrium does, but
in fact  the local generation of heat by thermonuclear processes creates
an outward flow of heat and is expected to explain the existence of a negative temperature gradient.
(It does, at least, explain the persistence of high temperatures in a system that is open to the cosmos.)

It is an indication of the incompleteness of this picture that it contains
no parameter that can be associated with the strength of radiation  and, {\it
a forteriori}, it does not allow us to investigate the  result of
turning off the radiation.

\b\b\b

\noÝ{\steptwo II. Thermodynamics}

\ce{\bf II.1. Thermodynamic equilibrium}

A state of thermodynamical equilibrium of a system that consists of a
very large number of identical particles is defined by the values of 3
variables, {\it a priori} independent, the volume $\V$, the pressure $P$
and the temperature $T$. These are variables taking real values;
they apply to the system as a whole. In the case of any particular system
there is one relation that holds for all equilibrium states, of the form
$$
T = f(\V,P).
$$
It is written in this form, rather than $F(T,\V,P) = 0$, because a unique
value of  $T$ is needed to define a state of equilibrium
between two systems that are in thermal contact with each other: it is
necessary and sufficient that they have the same temperature. This
statement incorporates the zeroth law.

 If we divide our system into subsystems then these will be in
thermal equilibrium with each other only if they  have the same temperature.
This, at least, is inherited wisdom.

The ideal gas at equilibrium is defined by global variables $T,\V,P$,  and two
relations. The principal one is the gas law
   $$
P\V ={{\cal R}\over \mu}T,~~ \R = .8314\times 10^{8} ergs/K,\eqno 
$$
where   $ \V$ is the volume of a mole
of gas. The other may take the form of an expression for the internal energy.
 
  \b\b

\ce{\bf II.2. The ideal gas in statistical mechanics}

Here
again we consider a gas that consists of identical
particles (Boltzmann statistics), each with mass $m$ and subject to
no forces. It is assumed that the $i$th particle has momentum $\vec
p_i$ and kinetic energy $\vec {p_i}^2/2m$.     It is assumed that the
number $N$ of particles with energy $E$ is given by the
Maxwell distribution
$$
N(E) \propto {\rm e}^{-E/kT},\eqno(2.1)
$$
which implies a
constant density in configuration space. Now place this gas in a
constant gravitational field, with potential $\phi(x,y,z) = gz,~ g$
constant. Since the potential varies extremely
slowly on the atomic scale it is plausible  that, at
equilibrium,  each horizontal layer ($\phi$ constant) is characterized by
a constant value of the temperature, density and pressure. Since
neighbouring layers are in thermal contact with each other the
temperature must (?) be the same throughout,
$$
T(z) = T = {\rm constant},\eqno(2.2)
$$
The energy of a particle at level $z$ is $\vec  p\, ^2/2m + mgz$ and (2.1)
now implies the following distribution in configuration space,
$$
\rho(x,y,z)  \propto {\rm e}^{-mgz/kT},\eqno(2.3)
$$
   in agreement with (1.17).
   This supports the expression for 
the potential, Eq.(1.15), which is strange since that potential is not 
appropriate for an ideal gas.
Both derivations of the distribution rest on the assumption that the
temperature is constant throughout the system.

 About the influence of gravitation on the temperature distribution there
has been some debate, see e.g. Waldram (1985), page 151. It is said that 
the kinetic energy of each atom in a monatomic gas is $3kT/2$ and that,
when the temperature is the same everywhere,  this is paradoxical because
it does not take account of the potential energy of the atom in the
gravitational field.    The incident involving Loschmidt, Maxwell and
Boltzmann has already been mentioned.   All
speculation along these lines falls short of being compelling. 
See the Kelvin footnote in Section I.7.

\b\b

\ce{\bf II.3. The first law and the internal energy}

  Can we extend the action principle to the case
that the temperature varies with time?
The action must be modified, for the temperature becomes
a dynamical field. Is the  temperature 
one of the variables with respect to which the action must be minimized?
  The
usual approach is to lay down the additional equation by fiat (Section
I.6); is this completely satisfactory? Would it perhaps be preferable to
have it appear as the result of minimizing the action with respect to
variations of the temperature field? We hope to show that there are
important advantages.

To prepare for the generalization we shall examine some of the
main tenets of thermodynamics in the context of the action principle.
Assume for the moment that the system is one of uniform density and
pressure.

Suppose  that the system is in  thermal and mechanical
isolation except for a force that is applied to the boundary. The system
is in an equilibrium state with temperature $T$. The applied force is
needed to hold the gas within the boundary of the domain $\Sigma$, then
decreased by a very small amount leading to a displacement of the
boundary and an increase of the volume by a small amount $d\V$. Assume
that this process is reversible. The work done by the applied force is
$$
dW = -p d\V.\eqno(2.4)
$$
The first law states that, if the system is in thermal isolation, then this
quantity is the differential of a function $U({\cal V}, S)$ that
is referred to
as the internal energy of the system.

Here it should be recalled that the natural variables of 
$U$ are $\V$ and $S$. 
When one says that the internal energy of an ideal gas is $n\R TM$,
what is meant is that this is the value obtained when $S$ is 
eliminated in favor of $T$ and $\V$.
Locally, as we shall see in Section II.7,
$$
u = n\R \rho T = (f+sT)|_{s = -{\p f /\p T}}.
$$
We regard $u$ as the on shell value of the function $V = f+sT$.

Consider  the system that consists of an ideal gas
confined to a volume
$\V$ and experiencing no external forces, not even gravitation. If the gas
expands at constant pressure the work done by the gas is $pd{\cal V}$ and
the ideal gas law Eq.(1.12) tells us that 
$$
pd\V = \R T\rho d{\cal V} = \R T {M\over {\cal V}}d {\cal V}. \eqno(2.5)
$$

The idea of energy conservation suggests a concept of ``internal
energy".
    It is assumed that, under certain
circumstances  ($dS = 0$), the work done by the gas  is at the expence of an internal
energy $U$ so that
$$
pd{\cal V} + dU = 0,
   $$
or
$$
{\cal R}TM d{\cal V}/{\cal V} + dU = 0. 
$$
It is  an experimental fact (Gay-Lussac 1827, Joule
1850) that the internal energy of an ideal gas is independent
   of the volume (see above) and the more precise statement
   that the on shell  internal energy density $u$ is proportional to ${\cal R}T\rho$
is often included in the definition of the ideal gas (Finkelstein 1969,
page 7). Thus
   $$
   u = \hat c_V{\cal R}T\rho,~~ U = \hat c_V {\cal R}TM.
   $$
Statistical  mechanics gives $\hat c_V = n$, where
   $n$ is the adiabatic index and takes the value $
3/2$ for a monatomic gas. Thus
   ${\cal R}TM d{\cal V}/{\cal V} + dU = {\cal R}TM d{\cal V}/{\cal V} + n{\cal R}MdT= 0$, which implies that
   $$
   dT = -{1\over n} {T\over {\cal V}} d{\cal V}, ~~T\propto  {\cal V}^{-1/n}. 
   $$
   The conclusion is that the two conditions that define an ideal gas, $p = \R \rho T$ and
   $u = n\R \rho T$,  imply the polytropic relation $\rho/T^n$ = constant for adiabatic changes.   
 The  calculation from  (2.4) onward was done with the understanding that $M = \rho\V$ is fixed.

At the deepest level the concept of energy
derives its importance from the fact that it is conserved with the
passage of time, by virtue of the dynamics.   In modern versions of
thermodynamics, and especially in the thermodynamics of irreversible
processes and in radiation thermodynamics, conservation laws are all
important, but they are postulated, one by one, not derived from basic
axioms as is the case in other branches of physics, and they have a 
formal aspect, serving to define various fluxes.
See e.g. Stanyukovich (1960), Castor (2004). 
  
\ve
 
\ce{\bf II.4. The first law and the hamiltonian}

{\it Having adopted an action principle approach we are bound to
associate the internal energy with the hamiltonian}.
 
The hamiltonian density is determined by the equations of motion only up
to the addition of a
constant multiple of the density. When we decide to  adopt a
particular expression to be used as internal energy over a range of
temperatures, we are introducing a new assumption.  Any expression for the
internal energy, together with the implication that applied forces increase
it by an amount determined by the work done, is a  statement about a family
of systems, indexed by the equilibrium temperature or the entropy. This cannot come out of the gas law
and implies an independent axiom.

If we adopt the simplest expression for the hamiltonian, that of the
isothermal atmosphere, with the lagrangian density of Eq.(1.16),
$$
H = \int d^3x  (\rho\vec v^2/2 + V), ~~ V = {\cal R}T\rho\ln\rho,
$$
 interpreting the potential as the `internal energy',  
then we shall get, in the static case,
$$
pd{\cal V} + dH(T,{\cal V}) = 0, ~~ p = {\cal R}TM/{\cal V}.
$$
with$$
dH = {\cal R}M\ln(M/ {\cal V}) dT -{\cal R}TMd{\cal V}/{\cal V}.
$$
The second term compensates for $pd{\cal V}$  and so $dT = 0$,
the temperature does not change. The temperature of
this gas is an adiabatic invariant.

   This  contradicts experimental results for   
ideal gases. In fact, the hamiltonian density $ h = V = {\cal
R}T\rho\ln\rho$ is not the correct expression for the internal energy
density of an ideal gas.   (We shall have more to say about this
later.)   

\b\b

\ce{\bf  II.5. The adiabatic  lagrangian}

In the absence of gravity the equilibrium configurations all have uniform
temperature, density and pressure. The equilibrium configurations described
by the lagrangian are  related by reversible
transformations involving no heat transfer, exactly the configurations
examined in the earliest experiments. In the presence of gravity  we
shall assume only that  the expression for the 
internal energy density has the same form, namely
$$
u = \hat c_V{\cal R}T\rho.\eqno(2.6)
$$

   Two kinds of additions can be made to the lagrangian (1.16)
   without spoiling the equations of motion that
   are essential to hydrodynamics.

Adding a term  linear in $\rho$ we
consider
$$
{\cal L}[\Phi, \rho,T] = \rho(\dot \Phi - \vec v\,^2/2 -\phi + \lambda )
   -{\cal R} T \rho \ln \rho + \rho\psi [T]  .\eqno(2.7)
$$
The continuity equation is unchanged, and the 
hydrodynamical equation remains
$$
\rho{D\over Dt}\vec v = -\rho\,{\rm grad}\,\phi - {\rm grad~}   p,~~ p = \R T\rho,
$$
where $p = \rho(\p V/\p \rho) - V$ is unchanged since the new term in the
potential is  linear in $\rho$ and since $\p V/\p T$ vanishes on shell.
(Section I.4.)

   Variation with respect to
$T$ gives
$$
\rho\psi'[T] - {\cal R}\rho\ln \rho  = 0.\eqno(2.8)
$$
On shell or, more precisely, by virtue of this equation, the potential reduces to
$$
V[\rho,T] = {\cal R}T\rho\ln \rho  -  \rho\psi [T] =
\rho \big(T\psi'[T]-\psi [T]\big)
$$
This is the on shell ``free", static hamiltonian density (gravitational
potential and kinetic energy omitted)  that we expect to identify with the
 internal energy density (2.6). This requires that 
$$
\psi [T]  =  n {\cal R} T\ln T/T_0 ,\eqno(2.9) 
$$
leaving free only the constant $T_0$.
The on shell hamiltonian  density takes the form 
$$
h| := h|_{\rm on~ shell} = \rho \vec v^2/2 + \rho\phi + u,~~ u = n\R T\rho. 
$$

Finally, the equation of motion (2.8), with $\psi [T]$ as  in (2.9), reduces
to
$$
  \R \rho(n-\log {k\over k_0}) = 0, ~~ k    = {\rho\over T^n}, ~~ k_0 = {\rm constant}.\eqno(2.10)  
$$ 
This is just the adiabatic relation $\rho/T^n$ = constant. It was derived, once more, from the two conditions that define an ideal gas.

The equation of motion that is obtained by variation with respect to
$\rho$ is
$$
\dot\Phi - \vec v^2/2 -\phi + \lambda    = {\cal
R}T\big(1+\ln {k\over k_0} \big). 
$$
Combined with Eq.(2.10) it reduces, in the static case, 
to
$$
 \phi-\lambda  +
(1+n)\R T   = 0,   
$$
which has the same form as the 
equation (1.18) studied in Section I.6, expressing the fact that the lapse
rate is constant. 

 {\it We have thus found an action principle, with dynamical variables
 $\rho$
and $T$, that   reproduces all of the equations that characterize the
equilibrium configurations,   as well as the
standard, hydrodynamical relations of an ideal gas.}

The lagrangian density is
$$
{\cal L}[\Phi, \rho,T] = \rho(\dot \Phi - \vec v\,^2/2 -\phi + \lambda )
   -\R T \rho \ln {\rho\over T^n k_0} .
$$

\ve

\ce{\bf II.6. The radiation term}

Since the advent of Schwarzschild's paper (1909), and especially after the
adoption of his ideas by Eddington (1926), the analysis of the effect of
radiation has not changed significantly. The conclusion, in the simplest
approximation, is that the internal energy is augmented by the
Stefan-Boltzmann expression for the energy of black body radiation. Thus
$$
{\cal L}[\Phi,\rho,T] = \rho(\dot\Phi - \vec v^2/2 -\phi + \lambda )
   -{\cal R}T\rho\ln  {k\over k_0} + {a\over 3} T^4 .\eqno(2.11) 
$$
The term $\rho \mu[T]$ in Eq.(2.7) has been included in the potential, 
with
$ 
k = \rho/T^n.
$ 
The constant $a = 7.64\times 10^{-15} ergs/K^4$ is the Stefan-Boltzmann
constant and the new term is the pressure of the photon gas, or of black
body radiation.

The new expression for the on shell internal energy is 
$$ 
u = n\R T\rho + aT^4 ,
$$
 and the pressure, defined either
by the modified hydrodynamical equation or by $dU + pdV = 0$, is
$$
p = \R T\rho + {a\over 3}T^4,
$$
in agreement with the theory of black body radiation and with the principle
that the pressure in a mixture of gases is additive. 

Variation with respect to $T$ now gives
$$
\R \Big(n- \ln {k\over k_0}\Big) \rho  +
{4a\over 3} T^3 = 0, \eqno(2.12)
$$
and in the important case when $n = 3$, 
$$
\R \Big(3 - \ln {k\over k_0}\Big){\rho\over T^3} +
{4a\over 3} = 0, 
$$
which is equivalent to Poisson's law $k = T^3/\rho = $ constant.
This reflects an affinity between the
polytropic ideal gas with $n = 3$  and radiation, strongly emphasized by Eddington (1926). The
value
$n = 3$ has a cosmological significance as well, it is characteristic of the changes
in $\rho, p, T$ induced by uniform expansion (Ritter , Emden 1907, see
Chandrasekhar 1938, page 48). For other values of $n$, Eq.(2.12) is a mild
modification of the polytropic equation of change in the presence of
radiation. The standard
approach maintains the polytropic relation without change in the
presence of radiation. 

The fact that the inclusion of the radiative term in the hamiltonian leads
to a modification of the polytropic relation illustrates  the new
perspective that characterizes the action principle formulation of
dynamics. 
It deserves to be emphasized that our approach does not allow us to
determine a relation between dynamical variables (such as $\rho \propto
T^n$ or even $P = \R T\rho$) until the lagrangian has been completed. The
modification of the polytropic relation that results from including the
energy of the radiation field is not a feature of the traditional theory.

 {\it We have thus found an action
that, varied with respect to $\rho, \Phi$ and $T$ 
   reproduces all of the equations that define the ideal,
polytropic gas with polytropic index $n = 3$, radiation included,  as well as
the standard, hydrodynamical relations.   For any  value of $n$, 
 it describes a gas that has its effective polytropic index
increased from the `natural' adiabatic value, approaching the critical value
3 at very high temperatures.}

 We suggest that
using the lagrangian (2.11) is preferable to the usual
assumption that 
$\beta := p_{\rm gas}/p_{\rm tot}$ is constant, which is true only when $n
= 3$. 

\b\b

\ce
{\bf II.7. Thermodynamical relations}

The basis for a close integration of the axtion principle with
thermodynamics is the identification of the hamiltonian with the function
$$
H = F(\V,T) + ST + P\V,
$$
a function of 4 independent variables $\V,T,S$ and $P$.

The states of thermodynamic equilibrium are the points in this
4-dimensional manifold at which $H$ is extremal with respect to
variations of $T$ and $\V$,
$$
{\p H\over \p T}\Big|_{\V, S, P} = {\p F\over \p T}\Big|_\V + S = 0,
$$
$$
{\p H\over \p \V}\Big|_{T, S, P}= {\p F\over \p \V}\big|_T +P = 0.
$$
In the case of an ideal gas,
$$
F(\V,T) = -\R T \ln (\V T^n),
$$
and these relations give
$$
S/\R =  n+\ln{ ( \V T^n)} ,
$$
and
$$ 
P\V = \R T.
$$
The on shell internal energy  density is
$$
U =\big(F(\V,Y) + ST\big)\Big|_{\rm on~ shell} = n\R T.
$$

There is a symmetry between energy and entropy, already emphasized by
Gibbs (1878). One can regard the entropy as a function of $\V,H$ and $P$,
$$
S = {1\over T}\big(H-F(\V,T) -P\V\big).
$$

\ve
The conditions that this function be stationary with respect to the
variables of $T$ and $\V$ are exactly the same,
$$
{\p S\over \p T}\Big|_{\V,H,P} = -{1\over T}\Big(S + {\p F\over \p
T}\Big|_\V \Big) = 0
$$
and
$$
{\p S\over \p \V}\Big|_{T,H,P} =  -{1\over T}\Big({\p F\over \p
\V}\Big|_T  + P\Big) = 0.
$$

The choice of the function $F(\V,T)$ characterized the system. The variable
$S$ remains a free parameter that is fixed when the system is isolated.

\b
\ce{\bf Local relations}

The local extrapolation of thermodynamics seeks to
promote these relations to field equations that describe local but stationary
configurations. The functions $F, H, S$ are given new interpretations as
specific densities, the variable $\V$ as specific volume. The mass
density is
$\rho = 1/\V$ and densities $f, h, s$ are defined by
$$
f(\rho,T) = \rho F(\V,T),~~ h = \rho H,~~ s = \rho S.
$$ 
The Hamiltonian is
$$
\int_\Sigma d^3x\,h,~~ h =   \rho \vec v^2\,/2 +
f(\rho,T) + sT  +P~.
$$

Variation with respect to $T$, with $\rho,S,T$ treated as independent
variables,  leads to
$$
{\p f\over \p T} + s = 0.\eqno(2.13)
$$
Variation with repsect to $\rho$, with the mass and the volume fixed,
 gives
$$
{h\over\rho} + \rho{\p (f/\rho)\over\p \rho} = \lambda. ~~(
\lambda  = {\rm constant}.)\eqno(2.14)
$$
The local thermodynamic  pressure is defined by 
$$
 \rho^2{\p (f/\rho)\over\p \rho}- p = 0,
$$
then the last relation  (2.14) reduces to
$$
{\rm grad}{h+p\over \rho} = 0,\eqno(2.15)
$$
 A complimentary variation of the density, with the mass
fixed but the volume not, gives the result that the internal pressure
$p$ must agree with the ``external" pressure $P$ on the boundary. This  is
the only way that the external pressure $P$ affects the equations of motion,
so the values assigned to $P$ in the interior play no role. However, one may
wish to assure that every part of the system is in equilibrium; that amounts
to placing imaginary boundaries;  then $P$ represents the pressure on the
boundary by the molecules that are on ``the outside", and then it has to
match the thermodynamic pressure $p$ everywhere. In this interpretation
$P$ is \underbar{the} pressure, an independent variable, and the condition
for equilibrium is that it take the value of the thermodynamic pressure $p$
everywhere.

In the case of the ideal gas 
$$
f(\rho,T) = \R\rho T\ln{\rho\over T^n}, ~~  p = \R \rho T.
$$ 
The on shell value of $h$ is the internal energy density,
$$
h|_{\rm on~ shell} =u =  \rho \vec v\,^2/2 + n\R \rho T 
$$
and the variational equations reduce  to
$$
S/\R  = n-\ln{\rho\over T^n},~~ {\rm grad} \Big(\vec v\,^2/2 + (n+1)\R T\Big)
= 0. 
$$

The first equation, (2.13), gives an expression for the entropy in terms of
the density and the temperature. It must  be interpreted as a condition to be
satisfied by
$\rho$ and $ T$ once the entropy $S$ is given. In Gibbs' original setting $S$
was just a parameter, here it is a distribution, but the interpretation is
the same.  Variation of the energy gives no information about the entropy
distribution. It retains its role as a free parameter, fixed for an isolated
system. 

The other equation, in the general case  Eq. (2.15), must be
compared with the Bernoulli equation. There is, however, an obstruction,
for Eq.(2.15) reads
$$
{\rm grad}(\vec v\,^2/2 + \phi + F + ST + {p\over\rho}) = 0,
$$
with the gravitational potential $\phi$
now included, while the Bernoulli equation is,
$$
{\rm grad}(\vec v\,^2/2 + \phi) + {1\over \rho}{\rm grad}\, p = 0.
$$ 
Now
$$
{\rm grad}\,(F + ST) = {\bf grad\,\rho}\,{\p\over \p\rho}(F +
ST) + T{\rm grad} S.
$$
The first term is ${\rm grad}\rho \,({p/ \rho^2}) = (1/\rho)\,{\rm grad}\,p -
 {\rm grad} \,(p/\rho)$, so that the  equation of motion is
$$
{\rm grad}(\vec v\,^2/2 + \phi) + {1\over\rho}{\rm grad} \,p + T{\rm grad}
\,S = 0.
$$
and 
the Bernoulli equation is obtained under the condition that the entropy is
uniform.

This, it turns out, is a very interesting requirement.
\b

 Reversing the roles of $h$ and
$s$ we extremize the total entropy
$$
\int_\Sigma d^3 x \,s,~~ s ={1\over T}\Big( h - \rho( \vec v\,^2/2+\phi) -
T{\rm grad}\, S f(\rho,T) -P\Big). 
$$
Variation with respect to $T$, with $\rho,T,P$ and $H = h/\rho$ treated as
independent variables,  gives the same result, ${\p f/ \p T} + s =0, $ but
variation with respect to $\rho$ gives an additional constraint, 
$$
S + {P-p\over \rho T} ={\rm constant}.\eqno(2.16)
$$
 This  result brings relief to  what is the greatest
difficulty of localized thermodynamics, finding the correct way to deal with
 the entropy. It also relates the two difficulties that arose in connections
with the equations of motion; the (inconsequential) one of fixing the role
of the external pressure, and the serious one of justifying the Bernoulli
equation, to each other.  In fact both problems are solved when we can show
that $P = p$, and in that case the entropy is uniform.

The result (2.16) is thus remarkble. The alternate form of Gibbs' axiom
actually provides exactly the additional information that is needed to
relate the axiom of maximum energy to the laws of hydrodynamics, provided
that the interpretation of $P$ is correct.

In the case of the earthly atmosphere, the implication is that, if the total
entropy is extremal, then the  entropy, and not the temperature, is uniform.

  If we accept   that the total entropy is
extremal with respect to variations of the temperature and the density,
then we must accept that the isolated gas takes on an equilibrium
configuration that resembles that of an actual, isentropic atmosphere.

In this section we have not invoked our action principle, only classical
ideas about the energy and the entropy being extremal at equilibrium. Let
us point out that the function that we referred to as the hamiltonian
actually deserves the name. We need to introduce a variable that is the
canonical conjugate to the variable $\rho$. The velocity potential
$\Phi$ fills the role admirably, with $\vec v = -{\rm grad}\, \Phi$ by
definition. The continuity equation takes the form
$ \dot\rho = -\d H/ \d \Phi$, the Bernoulli equation (in integrated form)
is $\dot\Phi = \d H/\d \rho$.

The variational principle is thus thoroughly integrated with
the thermodynamics of Gibbs.

\b\b

\ce{\bf A divided system} 

A well known general  calculation aims to show that, when two bodies are in
thermal contact, each in a state of equilibrium,  then
their temperatures will tend to equalize. Let us examine the case of an
ideal gas,
occupying both parts of a vessel that is divided by a weightless piston,
free to move in a vertical direction.

The total energy is a sum,
$$
H = \sum \int d^3 x\, h_i + P_i\V_i,~~ h_i = \rho(\vec v_i \,^2/2 + \phi) +
f_i + s_i T).
$$
Let us assume that the external potential $\phi$ is constant in each volume,
and that the masses are separately conserved.

Variations that hold the common boundary fixed lead to the usual expressions
for the entropy and the pressure, and a uniform entropy in each part. The
external pressures $P_1, P_2$ are not yet related to the thermodynamic
pressures $p_1, p_2$. It remains only to consider variations  of the form
$$
d\V_1 = -d\V_2 = -d\rho_1/\rho_1^2 = d\rho_2/\rho_2^2.
$$

Variation of the energy gives
$$
P_1-p_1   = P_2-p_2.
$$
It means that the difference between the thermodynamic pressures must
balance the net external force applied to the piston.

Variation of the total entropy gives
$$
S_1+{P_1-p_1  \over T_1} = S_2+{P_2-p_2  \over T_2}.
$$
The interpretation is this. If each subsystem is in equilibrium, $p_i =
P_i$, then the only requirement is that the entropies be equal. But if
the entropies are equal and both subsystems are out of equilibrium, then the
temperatures must be equal. That is, if $T_1\neq T_2$, then the system will
take advantage ot available thermodynamic changes to increase the entropy.

The usual argument rests on the reasonable assumption that, if the
temperatures are unequal, some process that may be related or not to the
considered thermodynamic changes, will intervene to modify the implications
of the first relation. However, it can perhaps be agreed that this is not a
prediction of thermodynamics in the narrow sense.

The conclusion that was reached above, that an equilibrium configuration of
an ideal gas is isentropic rather than isothermal, is derived from precisely
the same axioms. To modify the implications it is enough to invoke a
process, not yet taken into account, for the addition or withdrawal of energy
from a gas that is in equilibrium. The thermodynamics of an ideal gas  
cannot, by itself,  settle the question.

\b\b

\ce{\bf  II.8. The isothermal atmosphere}

In this section we study the isothermal atmosphere from the
conventional point of view.  Instead of the
polytropic relation, that theory postulates the equation of
continuity, the Bernoulli equation in differential form, and the 
``energy conservation equation"
 $$
\dot h + \vec \DD\cdot\big(\vec v\,(h+p)\big) = 0.
$$
 In the case of an ideal
gas in isolation, the familiar expressions $u =  n\R \rho T$ and $p = \R 
\rho T$ are assumed to hold. 

Given the other equations, the polytropic equation implies the energy
conservation equation, but is the inverse statement true?

We have
$$
\dot h = \dot \rho{\p h\over \p \rho} + \dot v{\p h\over \p v} + \dot T {\p h\over \p T}.
$$

\ve
In the first term we use the continuity equation,
$$
\dot \rho = -(\rho v)' = -({\p h\over \p v})' 
$$
to write it as 
$$
-{\p h\over \p \rho}({\p h\over \p v})' .
$$
The second term would compliment the first, 
to make a divergence,  but this is tricky. The Bernoulli equation gives
 $$
 -\dot v = (v^2/2)' + \phi' + {1\over \rho} p'.
 $$

Without any assumptions,   
$$
{1\over \rho} p' = {1\over \rho}(\rho \R T)' = {1\over \rho}(k \R T^{n+1})'
= (n+1)\R T' + {\R T\over k}k', ~~ k :=\rho/ T^n.
$$
Thus
$$
\dot h = -{\p h\over \p \rho}(\rho v)' - \rho v\cdot \Big(({h+p\over
\rho})' + {\R T\over k}k'\Big) + n\rho\R \dot T
$$
$$
= -\DD\dot\Big(v(h+p)\Big) + {\p p\over \p\rho} \DD(\rho v) -{\rho \R
T\over   k} v\cdot \DD k + n\rho \R T'.
$$
or
$$
\dot h + \vec \DD\cdot\big(\vec v\,(h+p)\big) =
-{\rho \R T\over k}(\dot k + \vec v \cdot \vec\DD k). 
$$
The right hand side can be transformed with the help of the equation of
continuity,$$
\dot h + \vec \DD\cdot\big(\vec v\,(h+p)\big) =   T(\dot s +
 \vec\DD \cdot \vec v s).
$$
where $s$ is the entropy density.\footnote *{If, besides being uniform,
the temperature is constant in time, then this equation expresses the
conservation of the free energy.  }

Conservation of entropy is a part of non equilibrium thermodynamics, 
but it is normally not invoked  in the case of an isolated system. The
equation $\dot s + \vec\DD\cdot (\s\vec v) = 0$ is, nevertheless, one of
the axioms of the general theory and it is natural to add it to the
equation of continuity and Bernoulli's equation as a basic equation of
the theory; for example, in the form
$$
\dot k + \vec v \cdot \vec\DD k = 0.\eqno(2.16)
$$
This makes energy conservation a theorem instead of an axiom. 
Even so, the injunction $T' = 0$
at equilibrium is an extra constraint that is not implied by 
the equations of motion.

What replaces or extends this rule in the case of  a non 
equilibrium situation? 
The heat equation comes to mind, but it does not imply 
that the temperature is 
constant at equilibrium, as  pointed out by Emden, {\it op cit}  
Section I.7.

Consider harmonic
perturbations of the isothermal equilibrium configuration, for which
$$
\rho'/\rho = k'/ k = -g/\R T.\eqno(2.17)
$$
Replace $\rho,T,k$ by $\rho + d\rho, T + d T, k + dk$ and suppose that
the space and time dependence of the  perturbations share the harmonic
factor
$
\e^{\alpha t + \beta z}, ~~  \alpha, \beta ~  complex~ constants.
$
Eq.(2.17) and the continuity equation give us
 $$
 {dk\over k}= {g\over\alpha T}v,~~ {d\rho\over \rho} = -{1\over \alpha}v' + {g\over \alpha T} v,~~
 {dT\over T} = -{1\over n\alpha}v'.
 $$
The Bernoulli equation is
$$
-\dot v = g + {1\over \rho}(\rho T)' = g + dT' + T{\rho'\over\rho}
$$
and after some calculation, 
$$
-\dot v = dT' + T({d\rho\over \rho})' = -{T\over n\alpha}v'' - {T\over
\alpha}v'' + {g\over\alpha} v'
$$
and
$$
\alpha^2 = \gamma\R T \beta^2 - g\beta.
$$

This formula has the familiar, laplacian limit in the absence of gravity,
$g\rightarrow 0$, and in the limit of high frequencies.
In terms of the wave number $\nu  = i\beta $
and the frequency $\omega =  \alpha/i$ the formula is
$$
\omega^2 = \gamma \R T\nu(\nu-{ig\over  \gamma \R T}).\eqno(2.18)
$$
If $\omega$ is real, which should be experimentally  
feasible, then
$$
 {\rm\,Im}~\nu = {g\over 2\gamma \R T},
$$
which makes the   factor $\exp (\beta z) =
\exp(-i\nu z)$ in the amplitude increase exponentially  with the altitude.
(Does it induce cooling?)

This invalidates the linear approximation at high altitudes, but the
predicted negative damping  is unexpected even in the case of a gas column of
modest height.

Finally,
$$
\omega^2 = \gamma\R T \Big(({\rm Re}\nu)^2 + ({g\over 2 \gamma \R T})^2 \Big).
$$
The bass section is silent.
 
Sound propagation in the polytropic atmosphere is unaffected by the
terrestrial gravitational field; the speed is the same in vertical and
horizontal directions. 
 
     The equations
that partially define the standard theory allow for an arbitrary fixed,
uniform value of the entropy. For any value of this number one
would obtain  a complete

\no and well defined adiabatic theory, but this would
violate the rule that $T$ must be uniform at equilibrium and is rejected. 
There remains only the three equations, including 
$\dot k + \vec v\cdot\vec\DD k = 0$. But the theory is incomplete, as is
evidenced by the fact that fixing the entropy is 
 still a mathematically
allowed option. The statement that the temperature must become uniform for
equilibrium configurations helps to pin it down, but it is still not
complete.

That the theory is incomplete is clearly shown by consideration of the
centrifuge.  The gas is moving and
the density is not constant. But the equivalence principle has been invoked,
to argue that  the temperature must be uniform in this situation as well. The
need to make this additional rule shows that, without it, the theory is
incomplete, nor does that make it complete. What is needed is a rule that
applies to non equilibrium configurations, from which the original rule of
uniform temperature can be inferrd by continuity.

 \b\b
\b

\ce{\bf II. 9. The puzzle} 

We return to our analysis of the isolated, ideal gas,  before the
introduction of the radiation term in Section II.6.  It was based on only
two assumptions, the familiar expression for the internal energy of an
ideal gas, and the ideal gas law. The adiabatic equation of change,
$\rho/T^n$ = constant, was not postulated but derived from those assumptions. 
In the absence
of gravity the result is standard. In particular, the equations of motion
allow for stationary states with uniform density and temperature. The effect
of gravity was included by adding the gravitational potential energy to the
hamiltonian; which is standard practice. The result of
that modification is that, in the presence of the gravitational field
there  are
no longer any stationary solutions with uniform temperature. Instead both
density and temperature decrease with elevation. This should be welcome as
being in agreement with what is observed  in real atmospheres.   The fact that a portion of the atmosphere of the Earth
exhibits the same temperature profile is a surprise; it suggests that the
temperature gradient is not a product of radiation

{\it The problem is that nothing could justify an application of this
theory to phenomena that are significantly influenced by radiation.}  
Nothing that went into building 
up the theory suggests that the gas is not in isolation.   But there is a very
strong conviction among physicists that, in an isolated system,  the
temperature must be uniform, gravitation notwithstanding. We have built a
theory that, surprisingly,  seems to apply to the irradiated  and gravitating
atmosphere, but we have not solved the  more basic problem, 
  to  provide a dynamical theory that incorporates the isothermal atmosphere.

 Why is the
prediction of a temperature gradient in an isolated atmosphere so shocking?
  Imagine a
large heat bath located in the region
$z>0$ in $\Rrm^3$. A vertical tube, filled with an ideal gas, has its upper
end in thermal contact with the bath, otherwise it is isolated. Assume that,
at equilibrium, the lower part of the tube has a temperature that is higher
than that of the bath.  Now extract a small amount of heat from the bottom
of the tube; then the restoration of equilibrium
demands that heat must flow from the bath to the warmer, lower part of the
tube,
   in
violation of one of the statements of the second law, namely: 
\b
``Heat cannot pass by itself from a colder to a hotter body"  (Clausius 1887).
\footnote *{ According to a recent experiment, it can! A recent preprint
(Graeff  2008) describes an experiment, carried out over a period of several
months, in which a persistent temperature gradient was observed in a column
of carefully isolated water.}
\b
Concerning the status of this formulation by Clausius of the second
law of thermodynamics we quote I. M\"uller (2007):
\b
``This statement, suggestive though it is, has often been criticised as
vague. And indeed, Clausius himself did not feel entirely satisfied with it.
Or else he would not have tried to make the sentence more rigorous in a
page-long comment, which, however, only  succeeds in removing whatever
suggestiveness the original statement may have had". And M\"uller continues:
"We need not go deeper into this because, after all, in the end there will
be an unequivocal {\it mathematical} statement of the second law".
\b

  We  note that Maxwell (1868), in
refuting Loschmidt, did not make use of the statement but  
argued   that the arrangement could be turned into a
source of energy, a second class perpetuum mobile. But the argument is
incomplete.

\b\b

 \ce {\bf II.10. The centrifuge and the atmosphere}

Kelvin justified the polytropic model of the atmosphere in 
terms of radiation
and convection. Eddington discounted the role of convection
 and relied on a
concept of radiative equilibrium.  To find out what happens 
\underbar{in the
case of complete isolation} we study the analogous situation in a
centerfuge. 

Consider an ideal gas. By a series of experiments in
which gravity does not play a role,  involving reversible
changes in temperature and pressure, it is found that, at equilibrium, the
laws $p/\rho = \R T$ and $\rho\propto T^n$ are satisfied, constant  $n$ 
fixed. When supplemented by the laws of hydrodynamics, they
are found to hold, or at least they are strongly believed to hold, in
configurations involving flow, over a limited time span, in the absence of
external forces.  
In addition it is believed that, at equilibrium, the temperature must
be  uniform. 
  Let us refer to this last statement as ``the rule".   
We have in mind a fixed quantity of gas contained in a vessel,
the walls of which present no friction and pass no heat.

Let the walls of the vessel be two vertical, concentric cylinders,
\footnote *{To avoid the objection that the velocity potential may not
exist as a one valued function in a non simply connected domain, we
divide the available volume into wedges. This shape reduces convection
and eliminates friction.} and  construct a stationary solution of the
equations of motion.   In terms of cylindrical coordinates, take
$v_z = v_r = 0, v_\theta =
\omega$, constant. The continuity
equation is satisfied with $\rho$ any function of $r$ alone. Then neither
$T$  nor
$p$  is constant, for the hydrodynamical equations demand  that
 $$
r\omega^2 = cT', ~~ c = (n+1)\R \approx  10^7 cm^2/sec^2 K~~
{\rm (for~ air)}.
$$ 
At first sight, this does not seem to violate the rule,  for this is not
a static configuration. But when the equivalence principle is brought to
bear, then it is argued, and we must concede, that the rule should be
generalized to cover this situation as well.

Instead, if we accept the foregoing calculations, and if we also accept
the equivalence principle,  then we shall be lead to expect that a
vertical column of an ideal gas, in mechanical equilibrium under the
influence of terrestrial gravity, \underbar{and perfectly isolated}, 
will have a pressure and temperature gradient exactly of the form
predicted by Homer Lane.  This  contradicts the
prevailing opinion of atmospheric scientists, that the temperature
gradient owes its existence to the heating associated with solar
radiation.  

Further measurements in the atmosphere may throw light on this,
but isolation is difficult. Experiments with a centrifuge
may be more realistic.  The temperature lapse rate is 
$r\omega^2\times 10^{-7} K/cm$. If the acceleration is 10 000 $g$ at the
outer wall, then the lapse rate will be 1 $K$ per 10 cm, using air.
The question of the existence of a temperature gradient is the most
urgent. Once this is resolved one way or another the approach to
equiilibrium is worthy of an investigation.

\b\b

\no{\steptwo III. Conclusions}

\ce{\bf  On variational principles}

The principal reasons for preferring an action principle formulation of 
thermodynamics were stressed in the introduction. Here we add some 
comments.

Variational principles have a very high reputation in most branches of
physics; they even occupy a central position in classical thermodynamics
of Gibbs (1878),
see for example the authoritative treatment by Callen (1960).
An action is available for the study of irrotational flows in hydrodynamics, see
e.g. Fetter and Walecka (1960), though it does not seem to have been much used.
Without the restriction to irrotational flows it remains possible to
formulate an action principle (Taub 1954, Bardeen  1970, Schutz  1970), but
the proliferation of velocity potentials is confusing and
no application is known to us.  

In this paper we rely on an action principle formulation of the 
laws that are believed to govern the behaviour of an ideal gas, in the
presence of gravity and radiation. We have restricted our
attention to irrotational  hydrodynamical  flows. With a hamiltonian
formulation there is a natural way to include the energy and pressure of
radiation, to cover the whole range from a dense gas at low temperatures
to a very dilute gas at the highest temperatures.   

It was shown that there is an action that incorporates all of the 
essential properties that characterize an ideal gas, expressed as
variational equations. The independent dynamical variables
are the density, the velocity potential and the temperature. The idea of
varying the action with respect to the temperature is unusual but much in the
classical tradition and the integration of the theory with classical
thermodynamics is complete (Section II.7).   The hamiltonian gives the
correct expression for the internal energy and the pressure.

It was shown,   in Section II.7, that the action principle is
fully integrated  and compatible with thermodynamics, in particular,
with Gibbs' extremum axioms of energy and entropy.

\b
\ce{\bf On isothermal atmospheres}

Into this framework the inclusion of a gravitational field is natural. 
Inevitably, it leads to pressure gradients and thus also temperature gradients.
The 
theory, as it stands, predicts the persistence of a temperature gradient in an
isolated system at equilibrium. The existence of a temperature gradient in an
isolated thermodynamical system is anathema to tradition, and further work is
required to find a way to avoid it, or to live with it. In the absence of
experimentation, the question may be said to be academic, for it has little
or no bearing on the application of the theory to actual atmospheres. But it
touches on the basics of thermodynamics and it deserves to be settled, or at
least debated. Here we shall try to summarize what it is that we feel is
missing in the official position.

Suppose we start with a vertical column that, for one reason or another, is
isentropic. At a certain moment we turn off the incoming radiation and
isolate the gas column from its environmant. Assume that the column
eventually becomes isothermal, and that the dissipation of the temperature 
gradient is a slow process during which the gas passes through a sequence of
adiabatic equilibrium configurations.  The question is this: what are those
intermeduary configurations? For example, does the polytropic index reach
the final value (infinity) continuously, or suddenly?

\b 

\ce{\bf The alternative}
The action principle advocated in this paper does not incorporate the heat
equation:
$$
A\dot T + \p_iC^{ij}\p_j T = 0,
$$
where $A$ is related to heat capacity and the tensor $C$ to conductivity.
 
Let us suppose that conductivity  and related phenomena that contribute to
$C$  are very small, so that the time  scale of diffusion is much longer
than that of the adiabatic excitations described by the equations of motion.
In normal atmospheres this is surely an excellent approximation. 
Then the heat equation will describe the slow cooling of the atmosphere that
will set in when radiation is switched off, and the system will pass through
a succession of configurations that we may take to be stationary and even
static.  
For this to happen the equations of motion must contain a variable that
parameterizes these configurations. Our lagrangian does in fact contain just
one free parameter.

The constant $k_0$ appears in the hamiltonian in the term
$$
-\R T\rho\log k_0,
$$
identified with entropy.
As we shall see (Appendix), this is very much like terms that appear when
the interaction of the gas with external fields  (electromagnetism) is taken
into account. The  interpretation must be that natural losses cause the decay
of $k_0$ and that the effect of the radiation that is incident on stable
atmospheres compensate for this loss.

We have not been able erase all doubt about the fundamental 
rule, that
  equilibrium implies a
uniform temperature in all cases.  Tolman (1934,
page 314) shows that, according to General Relativity, the  
temperature of an
isolated photon gas in a gravitational field is not quite uniform. 
The predicted
magnitude of this effect is very small, but it shows that there are
circumstances in which statistical mechanics is not the absolute truth.

The interaction of the ideal gas with electromagnetic fields has been
discussed in a provisional manner in the appendix. The transfer of
entropy between the two gases is in accord with the usual treatment of each
system separately. We found no  suggestion  that the interaction is
responsible for the temperature gradient.

\ve

\ce{\bf   Suggestions}

(1)   We propose the use of the
lagrangian (2.19), or its relativistic extension, with
$T$ treated as an independent dynamical variable and
   $n' = n$, in astrophysics . Variation with respect to $T$
yields  the adiabatic relations between $\rho$ and
$T$, so long as the pressure of radiation is negligible, but for
higher temperatures, when radiation becomes important, the effect is to 
increase
the effective value of $n'$ towards the ultimate limit 3, regardless of 
the
adiabatic index $n$ of the gas. See in this connection the discussion by
Cox and Giuli (1968),  page 271.

In the case that $n = 3$ there is Eddington's treatment of the
mixture of an ideal gas with the photon gas.  But most gas spheres have a
polytropic index somewhat less than 3 and in this case 
the ratio $\beta = p_{\rm gas}/p_{\rm tot}$ may not be constant throughout
the star. The lagrangian (2.11), with $n$ identified with the 
adiabatic index of the gas, gives all the equations that are used to
describe atmospheres, so long as radiation is insignificant.
With greater radiative pressure the polytropic index of the atmosphere is
affected. It is not quite constant, but  nearly so, and it approaches
the upper limit 3 when the radiation pressure becomes dominant. Eddington's
treatment was indicated because he used Tolman's approach to relativistic
thermodynamics, where there is room for only one density and only one
pressure.  Many kinds of mixtures have been studied, but the
equations that govern them do not supplement Tolman's gravitational concepts in
a natural  manner, in our opinion. Be that as it may, it is patent that the
approximation  
$\beta $ = constant, in the works of Eddington and Chandrasekhar, is a device
designed  to avoid dealing with two independent gases.

\b 
(2) In a subsequent paper we  make use of the  platform that
is provided by the action principle to study the stability of atmospheres.
A useful, exact virial thgeorem is derived from the  equations of motion. 
It is argued that expressions for the total energy are not enough to
determine stability; what is needed is an expression for the hamiltonian, in
terms of the dynamical variables of the theory. (Fronsdal 2009)

\b

(3) The theory has a natural extension to mixtures, incorporating the
Gibbs-Dalton hypothesis, It is being applied to mixed atmospheres, to
chemical reactions and to changes of phase. (Fronsdal
2010).

\b

(4) Observation of the diurnal and seasonal
variations of the equation of state of the troposphere may lead to a better
understanding of the role of radiation in our atmosphere.  The lapse rate
has been measured from the equator to the poles, summer and winter.  The
presence of  convection in the early hours of the day, known to gliders,
reduces the lapse rate but interpretation is difficult.  Observed reduction
by as much as a factor of 2 in the lapse rate  are also difficult to
interpret since the variation may be due, in part,  to unrecorded changes in
the composition of the atmosphere. What is needed is a survey of the
normalized lapse rate
$$
\tau := \mu {dT\over dz},
$$
where $\mu$ is the mean atomic weight. An observed relative constancy of
this parameter would support the expectations of the theory. 
 
The temperature lapse rate has been observed under conditions that would
not seem to favor it, between a snow covered surface and an inversion
layer. This is difficult to reconcile with the traditional point of view.

(5) The centrifuge
may be a more practical source of enlightenment. We understand that modern
centrifuges are capable of producing accelerations of up to 10$^6 g$. Any
positive result for the temperature gradient in an isolated gas would
 have important theoretical implications.

\ve

\no{\steptwo Appendix. Sources and entropy}
 
 \b

\ce{\bf A.1. Electromagnetic fields}  

We write the Maxwell lagrangian as follows,
$$
{\cal L}_{\rm rad} = {1\over 2\epsilon} \vec D^2 - {\mu\over 2}\vec H^2 +
 \vec D\cdot (\vec \p A_0 - \dot{\vec A}) -\vec H\cdot \vec \p\wedge\vec
A + JA,\eqno(A.1)
$$
and add it to the ideal gas lagrangian
$$
{\cal L}_{\rm gas} = \rho(\dot\Phi - \vec v^2/2 -\phi + \lambda )
 -{\cal R}T\rho\log k +    {a\over 3}T^4,
 \eqno(A.2)
$$
 Since  the susceptibility of
an ideal gas is small, the dielectric constant may be expressed by
$$
\epsilon = 1 +\kappa[\rho,T] , ~~{\rm or}~~ {1\over \epsilon} =
 1 - \kappa[\rho,T].\eqno(A.3)
$$
Paramagnetic effects will be ignored at present.
An interaction between the two systems occurs through the dependence of the
susceptibility on $\rho$. The source $S$ has become $-(\vec
D^2/2\rho)(\kappa/T)$. If this quantity has a constant value then it produces a
shift in the value of the  parameter $k_0$.

 Two interpretations are possible. The electromagnetic field may represent
an external field, produced mainly by the source $J$, and affecting the gas by
way of the coupling implied by the dependence of the dielectric constant on
$\rho$. Alternatively,   the field is produced by microscopic 
fluctuations, quantum vacuum fluctuations as well as effects of the
intrinsic dipoles of the molecules of the gas. In this latter case the main
effect of radiation is represented by the radiation term $aT^4/3$.  Our
difficulty is that neither interpretation is complete, and that we do not
have a sufficient grasp of the general case when either interpretation is
only half right. The following should therefore be regarded as tentative. 
 
Variation of the total action, with lagrangian ${\cal L}_{\rm rad} +
 {\cal L}_{\rm gas}$, with respect to $\vec A, \vec D, \vec H$ and $T$ gives
$$
\dot{\vec D} = \vec \p\wedge \vec H, \eqno(A.4)
$$
$$ 
\dot{\vec A} = \vec D/\epsilon,\eqno(A.5)
$$
$$
\mu\vec H = -\vec\p\wedge \vec A,\eqno(A.6)
$$
and
$$
\R(n-\log k)\rho -{\vec D\,^2\over 2}{\p\kappa\over \p T}+
 {4a\over 3}T^3=0.\eqno(A.7)
$$
Taking into account the first 3 equations we find for the static hamiltonian
$$
H =  \int d^3 x \Big( \phi\rho + \R \rho T \log k + {\vec D^2\over
2 } +{\mu \vec H^2\over 2}  -{\vec D^2\over 2}{ \kappa\over  T}-
{a\over 3}T^4 \Big).
$$ 
 With the help of (A.7) it becomes
 $$
 H = \int d^3 x\Big( \phi\rho + n\R \rho T + {\vec D^2\over 2 } +
 {\mu \vec H^2\over 2} + a T^4\Big) -\int d^3 x \,T{\vec D^2\over 2}{\p (T 
\kappa)\over \p T}.\eqno(A.8)
 $$
 The last term, from the point of view of  the thermodynamical
 interpretation of electrostatics, is 
 recognized as the entropy (Panofsky and Phillips 1955). On  a suitable 
choice of the functional
 $\kappa$ it merges into the 
 internal energy. For example, if $\kappa = \rho T$ it takes the form
 $\rho T S$ with $S = \vec D^2 $.
\b\b

\ce{\bf A.2.  Using  $T$ as a dynamical variable}

Let us examine the total lagrangian, 
$$
{\cal L} = {\cal L}_{\rm rad} + {\cal L}_{\rm gas} =
 \rho(\dot\Phi - \vec v\,^2/2 - \phi + \lambda) - \R T\rho\log{k\over k_0}
$$
$$
 + {\vec D^2\over 2\epsilon} + {\mu\over 2}\vec H^2 +
 \vec D\cdot (\vec \p A_0 - \dot{\vec A}) -\vec H\cdot \vec \p\wedge\vec A+
JA +{a\over 3}T^4.\eqno(A.9)
$$
So long as $\epsilon, \mu$ and $J$ are independent of $\rho,T$ and $\vec
v$, the variational equations of motion that are obtained by variation of
 $\vec v, \rho, \vec A, \vec H$ and $\vec D$ are all conventional, at least
when $n = 3$ (for  all $n$ if radiation is neglegible). It would be possible
to be content with that and fix
$T$ by fiat, as is usual;  in the case of the ideal gas without radiation the
result is the same.  But if $\epsilon$ depends on $\rho$ and  on $T$,
which is actually the case, then we get into a situation that provides the
strongest justification yet  for preferring an action principle formulation
with
$T$ as a dynamical variable. The equations of motion include a contribution
from the variation of $\epsilon$ with respect to $\rho$, so that one of the
basic hydrodynamical equations is modified. Thus it is clear that the
extension of the theory, to include the effect of radiation, is not just a
matter of including additional equations for the new degrees of freedom. The
presence of the term $\vec D^2/2\epsilon[\rho,T]$ certainly introduces the
density
$\rho$ into Maxwell's equations; that it introduces $\vec D$ into the
hydrodynamical equations is clear as well. {\it The over all
consistency of the total system of equations can probably be ensured by
heeding
 Onsager's principle of balance, but the action principle makes it
automatic.}

Variation of the action with respect to $T$ offers additional advantages.
The usual procedure, that amounts to fixing $ \rho = kT^n$, $k$ and $n$
constant, gives the same result when radiation is a relatively unimportant
companion to the ideal gas, but in the other limiting case, when the gas 
is very dilute and the material gas gas becomes an insignificant addition to
the photon gas, it is no longer tenable. We need an interpolation between the
two extreme cases and this is provided naturally by the postulate that the
action is stationary with respect to variations of the temperature field.

  In the absence of the ideal gas
we have another interesting system, the pure photon gas. The analogy 
between
the photon gas and the ideal gas is often stressed; there is an analogue 
of
the polytropic relation that fixes the temperature in terms of
$\rho$; the pressure of the photon field
 is   $(a/3) T^4$. Our lagrangian already contains this pressure; 
we should
like to  discover a closer connection between it and the electromagnetic
field.   In the limit when the density of the ideal gas is zero, Eq.(A.7)
becomes
$$
 -{\vec D\,^2\over 2}{\p\kappa\over \p T}+ {4a\over 3}T^3=0.  
$$
In the absence of the gas it is reasonable to impose Lorentz invariance, 
so
we include magnetic effects by completing the last to
$$
 -{ F\,^2\over 2}{\p\kappa\over \p T}+ {4a\over 3}T^3=0.  
$$
If we suppose that
$\kappa[\rho,T]$, in the limit
$\rho = 0$,
 takes the form  $\alpha T^2$, then
$$
 \alpha    F^2 = {4a\over 3}T^2.
$$
The radiation from a gas of Hertzian dipoles can be shown, with the help of
the Stefan-Boltzman law and Wien's displacement law, to satisfy a relation of
precisely this form. Whether the same relation holds in vacuum is uncertain,
but it is suggested by an analysis of the effective Born-Infeld lagrangian
calculated on the basis of the scattering of light by light (Euler 1936, 
Karplus and Neuman 1950). See also McKenna and Platzman (1962), Fronsdal
(2007).

 \b\b

\b\b 
\no{\steptwo Acknowledgements}

I thank R.J. Finkelstein, R.W. Huff, A. Kusenko, J. Rudnick, P.
Ventrinelli and G. Williams for discussions.

\ve
 
\no{\steptwo References}

\no Bardeen, J.M., A variational principle for rotating stars
in General Relativity, 

 Astrophys. J. {162}, 7 (1970).

\no Bernoulli, D., ~ Argentorat, 1738.

\no Boltzmann, L., Wissenschaftlidhe Abhandlungen, Hasenoehrl, Leipzig 1909. 

\no Callen, H.B., {\it Thermodynamics}, John Wiley N.Y. 1960. 

\no Carnot, S., quoted by Emden (1907).

\no Castor, J., {\it Radiation Hydrodynamics}, Cambridge U. press,  2004.
 
\no Chandrasekhar, S., {\it An Introduction to Stellar Structure}, U.
Chicago Press 1938.

\no Clausius, R., {\it Der mechanische W\"armetheorie"}, Vieweg Verlag,
Braunschwei 1887. 
 
\no Cox, J.P. and Giuli, R.T., {\it Principles of stellar structure}, Gordon
and Breach, 1968.

\no Eddington, A.S., {\it The internal constitution of stars}, Dover, N.Y.
1959
 
\no Emden,  {\it Gaskugeln}, Teubner 1907.

\no Euler, H.,    \"Uber die Streuung von Licht an Licht nach der
Diracschen Theorie,

 Ann.Phys.  {\bf 26} 398-?  (1936).

\no Fetter, A.L. and Walecka, J.D., {\it Theoretical Mechanics of Particles
and Continua},

\no Feynman, R.P., {\it Lecture Notes in Physics.} 
 
\no Finkelstein, R.J., {\it Thermodynamics and statistical physics}, W.H.
Freeman 1969.

\no Fronsdal, C., Reissner-Nordstrom and charged polytropes,

Lett.Math.Phys. {\bf 82}, 255-273 (2007). 

\no Fronsdal, C., Stability of polytropes, Phys.Rev.D, 104019 (2008). 

\no Fronsdal, C., ``Heat and Gravitation. II. Stability", ArXiv 0904.0427. 

\no  Fourier, J.B.J., {\it Th\'eorie analytique de la chaleur}, Didot, Paris
1822.

\no Gay-Lussac, J.L.,  ~The Expansion of Gases by Heat,  Annales de chimie
43 , 137- (1802). 

\no Gibbs, J.W., ``On the equilibrium of heterogeneous substances"
Trans.Conn.Acad.  1878. 

\no Graeff, R.W., Viewing the controversy Loschmidt-Boltzmann/Maxwell through

macroscopic measurements of the temperature gradients in vertical columns of
water, 

 preprint (2007). Additional  results are on the web
page: 'firstgravitymachine.com'. 

\no Joule, J.P.,  Remarks on the heat and constitution of elastic fluids,

Phil.Mag. IV 211 (1857).

\no Karplus, R. and Neuman, M.,
Non-linear Interactions between Electromagnetic Fields,

 Phys.Rev. {\bf 80} 380-385 (1950).

\no Kelvin, Thomson, W., Collected Mathematical and Physical papers, Vol. 5,
 232-235.

\no Kelvin, Thomson, W., Collected Mathematical and Physical papers, Vol. 3,
255-260. 

Cambridge U. Press 1911.
 
 \no Kelvin, Thompson, W.,  {\it Baltimore Lec tures},  C.J. Clay and Sons, London 1904.
 
\no Lane, H.J.,   On the Theoretical Temperature of the Sun, under the
Hypothesis of a gaseous 

Mass maintaining its Volume by its internal Heat,
and depending on the laws of gases 

as known to terrestrial Experiment, Amer.J.Sci.Arts, Series 2, {\bf 4}, 57- 
(1870).

\no Laplace, P.S., {\it Trait\'ee de Mechanique Celeste}, Paris 1825.

\no Loschmidt, L., Sitzungsb. Math.-Naturw.
Klasse Kais. Akad. Wissen. {\bf 73.2} 135 (1876).  

 \no Maxwell, J.C., The London, Edinburgh and Dublin Philosophical Magazine
{\bf35} 215 (1868).

\no McKenna, J. and Platzman,  P.M., Nonlinear
Interaction of Light in Vacuum,

  Phys. Rev. {\bf 129} 2354-2360 (1962).

\no M\"uller, I., {\it A History of Thermodynamics}, Springer, Berlin 2007. 

\no Panofsky W.K.H.  and Philips, M., {\it Classical Electricity and
Magnetism}, 

Addison-Wesley, Reading Mass.  1962.

\no Poisson, S.D., {\it Th\'eorie mathématique de la chaleur},  1835.

\no Ritter, A., A series of papers in Wiedemann Annalen, now Annalen der
Physik,

 For a list see Chandrasekhar (1938). The volumes 5-20 in
Wiedemann 

Annalen appear as the volumes 241-256 in Annalen der Physik.
 
\no Schutz, B.F. Jr., Perfect fluids in General Relativity: Velocity
potentials and a  variational 

principle, Phys.Rev.D {\bf 2}, 2762-2771 (1970). 

\no Schwarzschild, K., Ueber das Gleighgewicht der Sonnenatmosph\"are,

G\"ottinger Nachrichten,  41-53 (1906).

\no Stanyukovich, K.P., {\it Unsteady motion of continuous media},
Pergamon Press N. Y. 1960.
 


\no Taub, A.H., General relativistic variational principle for
perfect fluids, 

Phys.Rev. {\bf 94}, 1468 (1954).

\no Thomson, W., Lord Kelvin, On Homer Lane's problem of a spherical
gaseous nebula, 

Nature {\bf 75} 232-235 (1907).

\no Thomson, W., Lord Kelvin, On the convective equilibrium of temperature in
the 

atmosphere, Manchester Phil.Soc. {\bf 2} , 170-176 (1862). 

\no Tolman, R.C., {\it Relativity, Thermodynamics and Cosmology},
Clarendon, Oxford 1934.

\no Tolman, R.C., The electromotive force produced in solutions by
centrifugal action, 

Phys.Chem. MIT, {\bf 59}, 121-147 (1910).

\no Waldram, J.R., {\it The theory of electrodynamics}, Cambridge U. Press
1985. 
\end